\documentclass[ reprint,
 amsmath,amssymb,
 aps,
 prl
]{revtex4-2}
\usepackage[utf8]{inputenc}
\usepackage{amsmath}
\usepackage{color}
\usepackage{verbatim}
\usepackage{amsmath}
\usepackage{amssymb}
\usepackage{graphicx}
\usepackage{dirtytalk}
\usepackage{verbatim}
\usepackage{grffile}
\usepackage{dcolumn}% Align table columns on decimal point
\usepackage{bm}% bold math
\usepackage[dvipsnames]{xcolor}
\usepackage{siunitx}
\usepackage[normalem]{ulem}
\newcommand{\NaRb}{$^{23}$Na$^{87}$Rb}

\usepackage[colorlinks=true,allcolors=blue]{hyperref}% add hypertext capabilities

\begin{document}

\title{Observation of the Hanbury Brown and Twiss Effect with Ultracold Molecules}
\author{Jason S. Rosenberg}
\thanks{These authors contributed equally to this work.}
\affiliation{Department of Physics, Princeton University, Princeton, New Jersey 08544, USA}
\author{Lysander Christakis}
\thanks{These authors contributed equally to this work.}
\affiliation{Department of Physics, Princeton University, Princeton, New Jersey 08544, USA}
\author{Elmer Guardado-Sanchez}
\altaffiliation[Present address: ]{Department of Physics, Harvard University, Cambridge, MA 02138, USA}
\affiliation{Department of Physics, Princeton University, Princeton, New Jersey 08544, USA}
\author{Zoe Z. Yan}
\affiliation{Department of Physics, Princeton University, Princeton, New Jersey 08544, USA}
\author{Waseem S. Bakr}
\altaffiliation{Email: wbakr@princeton.edu}
\affiliation{Department of Physics, Princeton University, Princeton, New Jersey 08544, USA}

\date{\today}% It is always \today, today,

\begin{abstract}
Measuring the statistical correlations of individual quantum objects provides an excellent way to study complex quantum systems. Ultracold molecules represent a powerful platform for quantum science due to their rich and controllable internal degrees of freedom. However, the detection of correlations between single molecules in an ultracold gas has yet to be demonstrated. Here we observe the Hanbury Brown and Twiss effect in a gas of bosonic \NaRb, enabled by the realization of a quantum gas microscope for molecules. We detect the characteristic bunching correlations in the density fluctuations of a 2D molecular gas released from and subsequently recaptured in an optical lattice. The quantum gas microscope allows us to extract the positions of individual molecules with single-site resolution. As a result, we obtain a high-contrast two-molecule interference pattern with a visibility of $54(13)\%$. While these measured correlations arise purely from the quantum statistics of the molecules, the demonstrated capabilities pave the way toward site-resolved studies of interacting molecular gases in optical lattices.
\end{abstract}
\maketitle

In a landmark series of experiments in the 1950s, Hanbury Brown and Twiss (HBT) demonstrated bunching correlations of photons from chaotic sources of light arriving at two detectors~\cite{brown1954interferometer,brown1956correlation}. Their interferometry technique had practical applications in measuring the angular diameter of stars and led Glauber to develop a theory of quantum coherence, laying the foundation for the field of quantum optics~\cite{glauber1963quantum}. In contrast to conventional interference observed, for example, in Young's double-slit experiment, the HBT effect results from the interference of two-particle rather than single-particle amplitudes -- regardless of whether those ``particles" are photons, quasiparticles, or matter. HBT interferometry has become a workhorse in high energy and nuclear physics to probe the space-time geometry of collision volumes~\cite{baym1998hbt}. The effect has also been demonstrated with electrons~\cite{henny1999fermionic,oliver1999hanbury}, neutrons~\cite{iannuzzi2006direct}, and phonons~\cite{cohen2015phonon,riedinger2016non}.

In the field of ultracold quantum gases, the exquisite control afforded by modern experimental techniques has stimulated a wealth of intensity interferometry measurements in atomic systems, both bosonic~\cite{yasuda1996observation,folling2005spatial,ottl2005correlation,schellekens2005hanbury,jeltes2007comparison,hodgman2011direct, perrin2012hanbury,dall2013ideal,carcy2019momentum,tenart2021observation} and fermionic~\cite{rom2006free,jeltes2007comparison,preiss2019high}. Molecular gases have recently been brought into the ultracold regime where quantum effects play a significant role~\cite{bohn_cold_2017}, often by leveraging the powerful approach of associating two ultracold atoms into a single molecule~\cite{kohler_production_2006}. 
These molecules represent the most complex objects for which full control over all the motional and internal degrees of freedom has been demonstrated. Recent experiments with polar molecules in optical  lattices have studied strongly-correlated many-body systems~\cite{yan2013observation} and laid the groundwork for using molecules as qubits in quantum computing ~\cite{DeMille2002}.

\begin{figure}[h]
\includegraphics[width=\columnwidth]{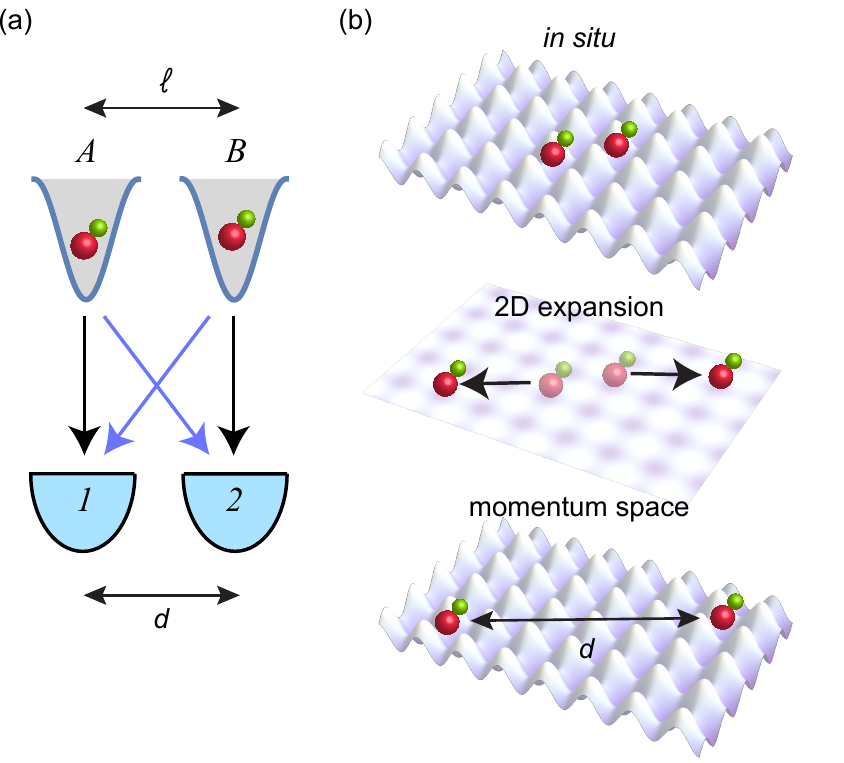}
\caption{\label{fig:1} Hanbury Brown and Twiss interference of heteronuclear bosonic molecules.  (a) Molecules released from lattice sites separated by a distance $\ell$ are detected after a time of flight at a variable separation \textit{d}. The amplitudes associated with the shown two-particle trajectories (black and blue) interfere, leading to a joint detection probability that depends on $d$. (b) The experiment in (a) is performed with molecules in an optical lattice. Top: multiple molecules are trapped in a 2D optical lattice. Middle: the 2D lattice is abruptly shut off, allowing the molecules to expand freely in the plane. An additional vertical lattice (not pictured) is left on to levitate the molecules against gravity. Bottom: the molecules are recaptured in the 2D lattice after time of flight, and their positions are measured with a quantum gas microscope.}
\end{figure}

\begin{figure*}[t]
\includegraphics[width=\textwidth]{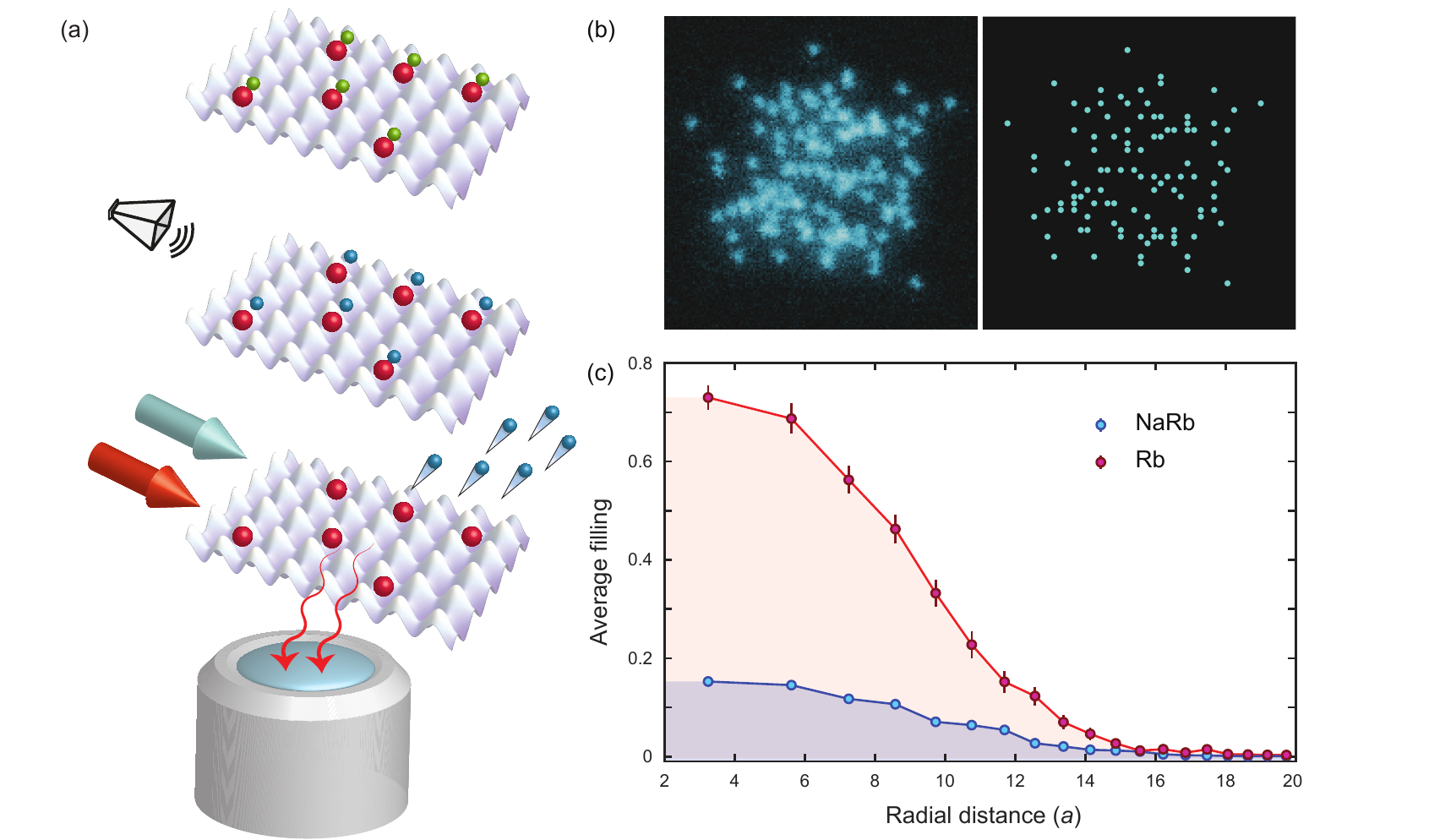}
\caption{\label{fig:2} Microscopy of molecules in an optical lattice. (a) Three-step process for detecting the molecules. Top: an array of NaRb molecules is prepared in a lattice. Middle: a microwave Landau-Zener sweep flips the Na atoms from the $\vert1,1\rangle$ to the $\vert2,2\rangle$ state, breaking apart the molecules into unbound Rb (red) and Na (blue) atoms. Bottom: a pulse of resonant light (blue beam) removes the Na atoms from the lattice. The remaining Rb atoms are laser cooled (red beam) and the fluorescence photons are collected with an objective, revealing the original positions of the molecules with single-site resolution. (b) Left: a sample fluorescence image of Rb atoms tagging the positions of molecules in the lattice. Right: molecule occupancy of the lattice for the fluorescence image. (c) Radially-averaged density profile of the NaRb molecules (blue) as well as the Rb atoms (red) prior to associating the atoms into molecules. We observe peak fillings of $0.73(2)$ for the Rb atoms and $0.15(1)$ for the molecules. Both density profiles are averaged over 30 experimental repetitions.} 
\end{figure*}
\begin{figure}[tb]
\includegraphics[width=\columnwidth]{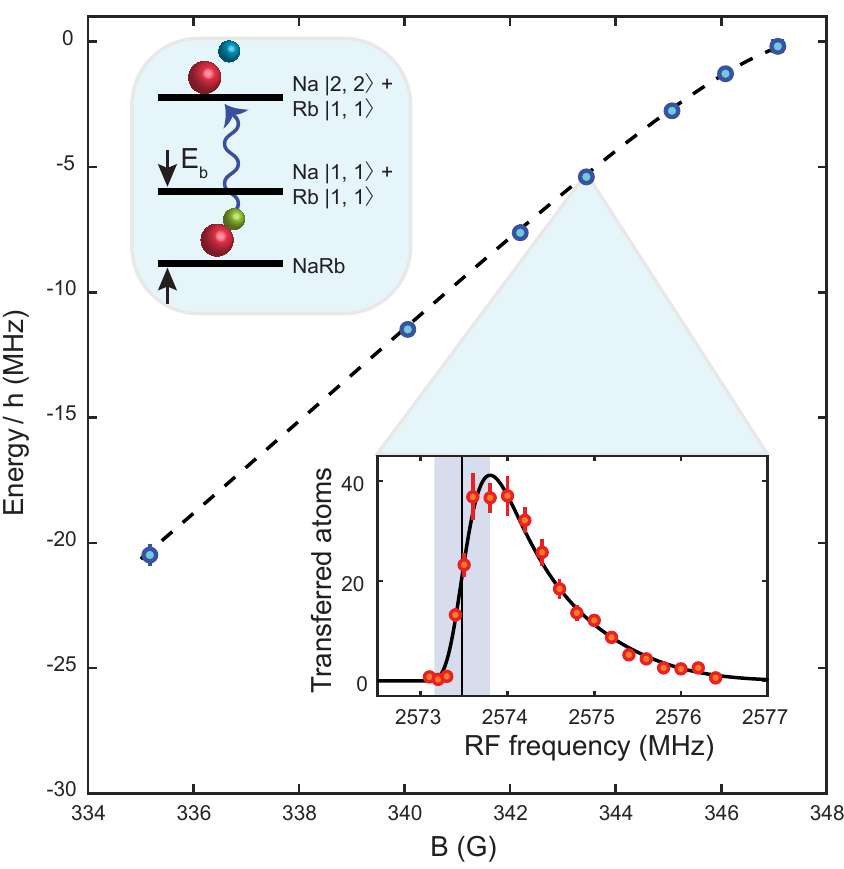}
\caption{\label{fig:3} Molecule binding energy versus magnetic field $B$. For each field, we drive the Na $|1,1\rangle{\rightarrow}|2,2\rangle$ transition to break apart the molecules (upper left inset) and measure the resulting dissociation spectrum. We extract the bound-free transition frequency from the onset of the spectrum. We also measure the free atomic transition frequency at each field. The difference between the atomic and molecular transition frequencies gives the binding energy (blue circles). The black dashed line is the predicted binding energy from a coupled-channel calculation. Lower right inset: example molecular dissociation spectrum at 343.4\,G (red circles) fit to an asymmetric Gaussian (solid black line). The error on the binding energies is taken as the half-width of the shaded region of the spectrum.}
\end{figure}

\begin{figure*}[tb]
\includegraphics[width=\textwidth]{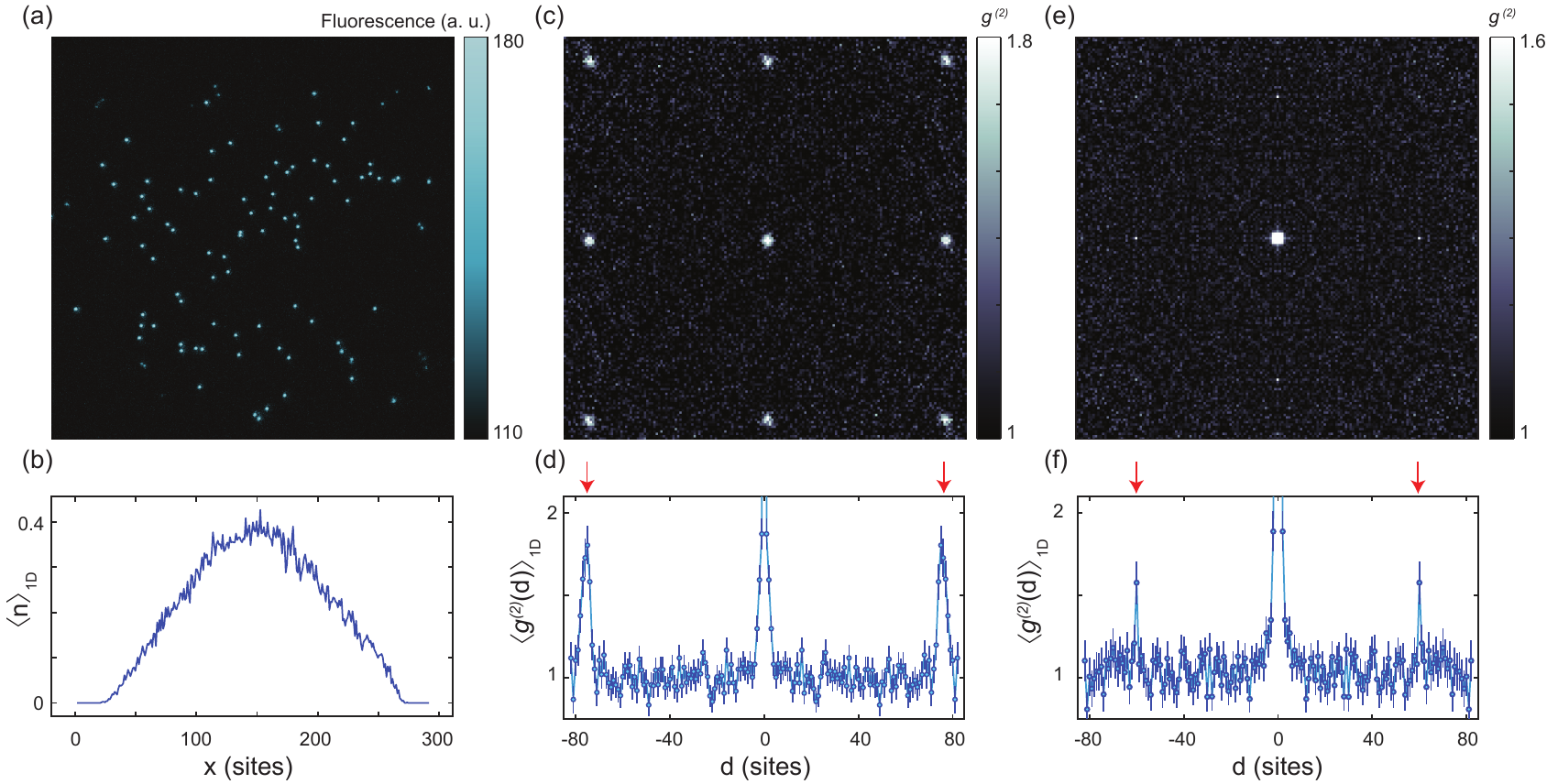}
\caption{\label{fig:4} Observation of the HBT effect with molecules. (a) Typical fluorescence image obtained after molecule time of flight. (b) Integrated density profile after time of flight, averaged over 764 images. (c) HBT correlations for Rb atoms, post-selected for images that yielded between 130 and 300 atoms. (d) 1D cut of the Rb HBT correlations averaged along the horizontal and vertical axes. Red arrows indicate the predicted position of the interference peaks. (e) HBT correlations for molecules averaged across the lattice symmetries. Images with fewer than 20 molecules were excluded from the analysis. (f) 1D cut of the NaRb HBT correlations averaged along the horizontal and vertical axes. Red arrows indicate the predicted position of the interference peaks.}
\end{figure*}

As two-particle correlations will form a crucial experimental probe for interacting many-body systems comprising polar molecules, we first study whether intensity interferometry can be used to characterize distributions of non-interacting molecules. We perform an HBT experiment in which we measure fluctuations in the density distribution of individual clouds of molecules on a grid of detectors after a time of flight~\cite{altman2004probing}. Particles originating from sources $A$ and $B$ can arrive at detectors $1$ and $2$ in two different ways ($A\rightarrow 1, B\rightarrow 2$ and $A\rightarrow 2, B\rightarrow 1$), shown in Fig.~\ref{fig:1}. If the two paths are indistinguishable then their quantum mechanical amplitudes interfere, as reflected in the two-point density correlation function.  The correlation relates to the Fourier transform of the source density distribution in the case of a distribution of sources with no phase coherence. As a result, if the positions of the molecules are initially discretized by an optical lattice, the interference pattern exhibits constructive interference peaks whose separation and width are inversely related to the lattice spacing and $\textit{in-situ}$ cloud size, respectively~\cite{folling2005spatial,rom2006free}.

There are two requirements for measuring a high-visibility HBT interference pattern: particle shot noise-limited detection and a detector spatial resolution better than the interference peak width. To fulfill these conditions, we detect the positions of individual molecules with single-site resolution using a quantum gas microscope~\cite{bakr2009quantum,gross2020quantum}. To date, single neutral ultracold molecules have only been detected in small tweezer arrays containing up to four molecules~\cite{anderegg2019optical,zhang2020forming,he2020coherently}. With the molecular quantum gas microscope introduced in this work, we prepare and detect over one hundred molecules simultaneously by dissociating them into atoms and using the atoms as tags for the molecule positions.

We form ultracold bosonic \NaRb~molecules from degenerate gases of Na and Rb atoms by magnetoassociation~\cite{wang2015formation}. 
To start, dual Na and Rb Bose-Einstein condensates (BECs) are prepared in a single layer of a vertical lattice. Fine adjustment of atom numbers is accomplished using a bichromatic dimple trap, allowing the reproducible production of clouds of several hundred atoms per species. Both species are prepared in their $\left|F=1,m_F=1\right\rangle$ states, where $F$ and $m_F$ are the total angular momentum quantum number and its projection along the quantization axis, respectively. Next, we make the mixture miscible by nulling the interspecies scattering length using an $s$-wave Feshbach resonance at 347.6\,G~\cite{wang2015double}. The atoms are then loaded into a 2D in-plane optical lattice with spacing $a\,{=}\,752$\,nm, and tunneling between sites is frozen
by increasing the lattice depth. We associate the atoms into molecules by adiabatically ramping the magnetic field across the resonance at a rate of 2.5\,G/ms, changing the scattering length from attractive to repulsive. For lattice sites containing one atom of each species, the theoretical conversion efficiency is close to unity (Supplementary Information). We selectively remove remaining free Na and Rb atoms by transferring them to the $\vert2,2\rangle$ hyperfine state and applying resonant light pulses on the optical cycling transitions. The transfer and removal is repeated four times for each species to achieve high efficiency.

To realize a quantum gas microscope for molecules, we employ a three-step process to extract site-resolved positions, as shown in Fig.~\hyperref[fig:2]{2(a)}. We first ramp the magnetic field back to the attractive side of the Feshbach resonance, increasing the Franck-Condon overlap between the bound and free states of atoms on a site. We note that a confinement-induced weakly-bound state exists here due to the lattice~\cite{stoferle2006molecules}, so an adiabatic ramp across the resonance does not break apart the molecules. The confinement-induced molecules are then dissociated by addressing Na with another series of microwave transfers and optical removal pulses. The remaining Rb atoms serve as markers for the Feshbach molecule locations. Light-assisted collisions during the optical removal pulses are estimated to be negligible (Supplementary Information). After ramping the magnetic field to zero, the in-plane lattice depth is increased to 6000\,$E_{\text{r,Rb}}$ (where $E_{\text{r,Rb}}=h^2/8m_{\rm Rb}a^2$ is the recoil energy for Rb) and the atoms are loaded into a light sheet potential. Optical molasses beams both cool the Rb atoms and scatter photons, which are collected with a high-numerical-aperture objective and imaged onto a camera. The full width at half maximum of the point spread function is $1010(15)$\,nm, which is sufficient for single-site resolution using a reconstruction algorithm~\cite{gross2020quantum}.

Fig.~\hyperref[fig:2]{2(b)} shows an example image of an \textit{in-situ} cloud of 103 molecules. A radially-averaged density profile, averaged over 30 experimental repetitions, is shown in Fig.~\hyperref[fig:2]{2(c)}. We measure a central molecule filling of $0.15(1)$, compared to the Rb parity-projected central filling of $0.73(2)$. The molecule filling fraction can be increased in future work by improving the overlap between the atomic clouds, lowering their temperature, and increasing the vertical confinement to obtain stronger interactions at the superfluid to Mott insulator transition.

A high molecule detection fidelity is essential for obtaining high contrast HBT interference as well as for future applications of the molecule microscope. Possible limitations on the molecule detection fidelity include imperfect atom-tagging of sites that had molecules as well as site-to-site hopping and loss of the Rb tag atoms during fluorescence imaging. We first measure the false negative molecule detection rate due to imperfect tagging. After performing the tagging protocol consisting of dissociating the molecules and optically removing Na, we ramp the field to the repulsive side of the Feshbach resonance and remove the Rb tag atoms. The false negative rate of $1.2(1)$\% is obtained by repeating the tagging protocol prior to fluorescence imaging to identify failures from the first attempt. During our 0.5\,s imaging exposure time, we measure site-to-site hopping of $0.1(1)$\% and $1.7(3)$\% loss. The latter is consistent with the measured atom lifetime limited by background gas collisions. We additionally measure a false positive rate of $0.34(5)$\% due to Rb atoms that did not associate into molecules and were not removed before imaging (Supplementary Information).

We further confirm that the detected particles are molecules by measuring their binding energy as a function of magnetic field using dissociation spectroscopy (Fig.~\ref{fig:3}). Driving the Na $|1,1\rangle{\rightarrow}|2,2\rangle$ transition yields the molecular dissociation spectrum, showing a sharp onset at a microwave frequency shifted by the binding energy $E_{\rm b}$ from the atomic transition.
Molecules that are not dissociated are not detected in the fluorescence imaging. 
We find close agreement between our measured binding energies and a coupled-channel calculation using the \textsc{bound} package~\cite{hutson2019bound} based on the parameters in Ref.~\cite{guo2021tunable}. Confinement effects are ignored since they are smaller than the resolution of the measurement.

Having established our molecule detection procedure, we now observe the HBT effect by measuring the density-density correlation function $g^{(2)}(\textbf{d})$ after a long time of flight (TOF) (Fig. \ref{fig:4}). The correlation function is defined as:
\begin{equation}
g^{(2)}(\textbf{d}) = \frac{\int \langle \hat{n}(\textbf{x}) \hat{n}(\textbf{x+d}) \rangle d^2 \textbf{x}}{\int \langle \hat{n}(\textbf{x}) \rangle \langle \hat{n}(\textbf{x+d}) \rangle d^2 \textbf{x}} 
\end{equation}
where $\hat{n}(\textbf{x})$ is the number operator at detection position $\textbf{x}$, and $\textbf{d}$ is the displacement between the detection positions. Maximal particle bunching is indicated by $g^{(2)}\,{=}\,2$ while for uncorrelated particles $g^{(2)}\,{=}\,1$. For a Gaussian source cloud of half-width $s$ (at $e^{-1/2}$ of maximum), the half-width of the interference peaks is given by $\delta\,{=}\,\hbar t/ms$ (at $e^{-1}$ of maximum), where $m$ is the particle mass and $t$ is the TOF. While particle shot noise-limited imaging in a 2D plane can theoretically achieve peak correlation amplitudes of $g^{(2)}\,{=}\,2$, the amplitude will be reduced if the width of the peaks is narrower than the detector size, in our case one lattice site. Therefore, we carefully choose our source cloud size and use the largest possible TOF given the constraints from the size of the lattice beams to maximize the HBT amplitude. 

A larger signal-to-noise ratio of the interference pattern can be achieved for higher \textit{in-situ} lattice filling fractions. Since we achieve higher fillings with atoms than with molecules, we first benchmark the interferometry protocol with Rb atoms. We prepare a gas with $189(20)$ Rb atoms frozen in a 2D optical lattice (66 $E_{\text{r,Rb}}$ depth), with a peak filling of $0.86(2)$ and an average source size $s\,{=}\,7(1)$ sites. We abruptly turn off the 2D lattice to initiate a $9.4(1)$\,ms TOF in the vertical lattice. The vertical lattice confinement is set to $\omega_\text{Rb} \,{=}\, 2\pi\times (3,4,1000)$\,Hz, providing negligible radial confinement.  
The TOF satisfies the far-field detection condition $t \gg 2\pi/\omega_{r}$, where $\omega_{r}\,{=}\, 2\pi\times 14$\,kHz is the on-site radial trap frequency of the 2D lattice for Rb. Following the TOF, we turn on the 2D lattice to pin the distribution for imaging.

The observed atomic HBT correlations are shown in Fig.~\hyperref[fig:4]{4(c)}. We observe a high contrast interference pattern with average correlation peak amplitudes of $1.80(12)$ and an average background value of $0.999(6)$. The measured peak separation and width from the 1D cut in Fig.~\hyperref[fig:4]{4(d)} is $75.9(4)$ and $2.4(2)$ sites respectively, close to the theoretically expected spacing of $h t/m_{\rm Rb} a \,{=}\, 76.2(8)$ and $\delta\,{=}\,1.7(3)$ sites. The symmetric pattern verifies that the 2D lattice axes are orthogonal and the lattice spacings along both axes are identical to better than $0.5$\%. 
This implies that $g^{(2)}(\textbf{d})$ is invariant for $\textbf{d}$ reflected across the $x\,{=}\,0$, $y\,{=}\,0$, $x\,{=}\,y$, and $x\,{=}\,-y$ symmetry axes, justifying averaging of the weaker molecular correlations across the lattice symmetries to reduce noise. 

We repeat the HBT interferometry with 56(13) molecules and a mean source size $s\approx\,17$ sites, which is expected to produce interference peaks whose widths are on the order of the lattice spacing. The protocol is the same as that used for atoms, with the molecules released from the 2D lattice at a magnetic field of $335.1$~G (with binding energy $E_\text{b}/h\approx 20$\,MHz). Fig.~\hyperref[fig:4]{4(e,f)} shows the observed molecular correlations averaged across the lattice symmetries. Since the TOF is the same as that used in the Rb atom correlation measurement, the smaller correlation peak spacing for the molecules is a direct result of their increased mass. The measured spacing is $60.0(5)$ sites, consistent with the theoretical expectation of $60.3(6)$ sites. While the correlation peaks are narrower than for Rb ($<1$ site), the peak amplitude remains large at 1.58(13). The average baseline is 1.04(1), with the deviation from unity caused by correlations at all distances due to shot-to-shot molecule number fluctuations. The interference contrast is sensitive to the preparation of the molecules in the same internal state and the same motional state of the vertical lattice, since these quantum numbers can provide which-path information during the free expansion; therefore, the measured contrast of $0.54(13)$ indicates a high degree of indistinguishability of the molecules. 

To conclude, we have demonstrated site-resolved measurements of density correlation functions in a non-interacting molecular quantum gas after time of flight expansion. The observation of the HBT effect with molecules paves the way toward realizing other quantum optical phenomena with molecules of increasing complexity~\cite{mitra2020direct}. By transferring the Feshbach molecules to the rovibrational ground state~\cite{guo2016creation}, a molecular lattice gas with strong dipolar interactions can be prepared~\cite{bohn_cold_2017}.
The interplay of quantum statistics and interactions can give rise to interesting signatures in HBT measurements~\cite{carcy2019momentum,tenart2021observation}. In addition, many correlated quantum states predicted to be realizable with polar molecules also exhibit real-space density correlations that can be directly measured with a molecule microscope. These include Wigner crystals~\cite{buchler2007strongly} and Mott solids with rational lattice fillings~\cite{capogrosso2010quantum}. Finally, inelastic collisions in molecular gases can lead to strong correlations through the quantum Zeno effect. While previous experiments have observed these correlations through the inhibition of loss~\cite{syassen2008strong,yan2013observation}, a molecule microscope would enable their spatially-resolved measurement.

We would like to thank Geoffrey Zheng, Sejal Aggarwal, Alan Morningstar, and Ravin Raj for experimental assistance. This work was supported by the NSF (grant no. 1912154) and the David and Lucile Packard Foundation (grant no. 2016-65128). L.C. was supported by the NSF Graduate Research Fellowship Program.
\bibliography{molecule_microscope}

\section{Methods}
Here we describe the formation of ultracold bosonic \NaRb~molecules by magnetoassociation from degenerate gases of Na and Rb atoms. NaRb molecules have been produced in bulk mixtures in Ref.~\cite{wang2015formation}. 

We start by creating dual Na/Rb Bose-Einstein condensates (BECs) with typically $2 \times 10^5$ atoms of each species in the $\left|F=1,m_F=-1\right\rangle$ state, using Na as a sympathetic coolant for Rb during forced evaporative cooling in a quadrupole magnetic trap followed by evaporation in a crossed 1064\,nm optical dipole trap. To prepare a 2D system, we load the condensates into a 1064\,nm light sheet with tight confinement along the vertical direction ($\omega_\text{Na} = 2\pi\times (24,117,1900)$~Hz, $\omega_\text{Rb}=0.86\,\omega_\text{Na}$). We then transfer both atom species to the $\left|F=1,m_F=1\right\rangle$ state, the entrance channel for the relevant Feshbach resonance. To further increase the vertical confinement, the atoms are loaded into a single layer of a 3.8\,$\mu$m spacing vertical lattice created by two 1064\,nm beams intersecting at 16$^\circ$ ($\omega_\text{Na} = 2\pi\times (14,20,4500)$\,Hz). For our optical lattice approach of molecule formation, we need the central density of the clouds to be on the order of one atom per site for each species. Given our trap parameters, this requires the ability to reproducibly generate small condensates of order one hundred atoms. We achieve this by performing a second stage of evaporative cooling in a tightly focused bichromatic dimple trap, which allows for independent adjustment of the atom number in each species (Supplementary Information).

Near zero magnetic field, the Na and Rb BECs are immiscible. To increase the overlap of the spatial distributions prior to magnetoassociation, we tune the interspecies scattering length using an $s$-wave Feshbach resonance at 347.6\,G to bring the clouds into a miscible regime~\cite{wang2015double}. We quickly ramp the field above the resonance to 415.9\,G, then slowly decrease the field to the zero crossing of the interspecies scattering length at 351.9\,G in 20\,ms. We subsequently load the 2D mixture into an in-plane square lattice with spacing $a=752$\,nm created by the fourfold interference of a single 1064\,nm beam (105\,$\mu$m waist) in a bowtie configuration. We freeze tunneling for both species by ramping the in-plane lattice depth to 36\,$E_{\text{r,Na}}$, where $E_{\text{r,Na}}=h^2/8m_{\rm Na}a^2$ is the recoil energy for Na. Next, we ramp the magnetic field below the resonance at a rate of 2.5\,G/ms to form Feshbach molecules. For lattice sites containing one atom of each species, the theoretical conversion efficiency is very close to unity (Supplementary Information). We selectively remove remaining free atoms by transferring them to $\vert2,2\rangle$ with a microwave Landau-Zener sweep at a field of $\sim 346.6$\,G (molecule binding energy $E_\text{b}/h\approx 0.7$\,MHz) and applying resonant light on the $\vert2,2\rangle$ to $\vert3,3\rangle$ optical cycling transition. The transfer and removal is repeated four times for each species to achieve high efficiency as described in the main text.

For imaging, the magnetic field is brought to 0\,G, the in-plane lattice depth is increased to 6000\,$E_{\text{r,Rb}}$ and the Rb atoms are loaded back into the light sheet at a depth of 140\,$\mu$K. Optical molasses cooling light scatters photons into a 0.5~numerical aperture objective, with $\sim10^4$ photons collected on the camera per atom per second.

%%%%%%%%%%%
\setcounter{figure}{0}
\setcounter{equation}{0}
\setcounter{section}{0}

\clearpage
\onecolumngrid
\vspace{\columnsep}

\newcolumntype{Y}{>{\centering\arraybackslash}X}
\newcolumntype{Z}{>{\raggedleft\arraybackslash}X}

\newlength{\figwidth}
\setlength{\figwidth}{0.45\textwidth}

\renewcommand{\thefigure}{S\arabic{figure}}
\renewcommand{\theHfigure}{Supplement.\thefigure}
\renewcommand{\theequation}{S\arabic{equation}}
\renewcommand{\thesection}{\arabic{section}}

\begin{center}
	\large{\textbf{Supplementary Information}}\\~\\
	
	\normalsize{Jason S. Rosenberg, Lysander Christakis, Elmer Guardado-Sanchez, Zoe Z. Yan, and Waseem S. Bakr\\
    \textit{Department of Physics, Princeton University, Princeton, New Jersey 08544, USA}
    
}
\end{center}

\section{Preparation of a dual 2D BEC}
Our experiment begins by loading a dual-species magneto-optical trap (MOT) with typically $8 \times 10^8$~Na atoms and $3\times 10^6$~Rb atoms from two separate 2D-MOTs in 8\,s. We deliberately load fewer Rb atoms than Na because we use the Na as a sympathetic coolant for the Rb during forced evaporative cooling. For Na, we employ a dark-spot MOT to increase its initial phase space density. We then cool the Na atoms in an optical molasses, at the end of which the temperature of the Na atoms is $\sim140$\,$\mu$K. The $\left|F=1,m_F=-1\right\rangle$ atoms are loaded into a magnetic quadrupole trap by ramping a magnetic field gradient to 174\,G/cm. The temperature of the mixture after loading the magnetic trap is $\sim 160\,\mu$K. We perform forced microwave evaporation of the Na atoms by flipping atoms in $\left|1,-1\right\rangle$ to $\left|2,-2\right\rangle$. The Rb atoms are sympathetically cooled. In order to minimize atom loss near the center of the trap where the magnetic field goes to zero, we plug the trap with a repulsive laser beam using 10\,W of 532\,nm light focused to a 45\,$\mu$m waist. 

The evaporation in the magnetic trap lasts for 20\,s. Halfway through the evaporation, a crossed optical dipole trap (XODT) is ramped up to 18\,W. The dipole trap is created using a single beam at 1064\,nm folded into a $90^\circ$ bowtie configuration, with the two intersecting arms orthogonally polarized. The waist at the atoms is 105\,$\mu$m. At the end of the evaporation in the quadrupole trap, the magnetic field is ramped down and the atoms are transferred to the optical trap. At this point, the atom number is $2.3\times10^6$ ($6\times10^5$) for Na (Rb) and the temperature of the mixture is 10\,$\mu$K. We perform an optical evaporation by lowering the depth of the optical trap exponentially to 340\,mW over 4\,s. A bias magnetic field of 8\,G is applied during the optical evaporation to suppress spin-changing collisions. To avoid differential gravitational sag between the Na and Rb atoms, we turn on an 80\,mW 1064\,nm light sheet beam (waists $w_z=14$\,$\mu$m, $w_r=235$\,$\mu$m) during the XODT evaporation.

\begin{figure}[h]
\includegraphics[width=.5\columnwidth]{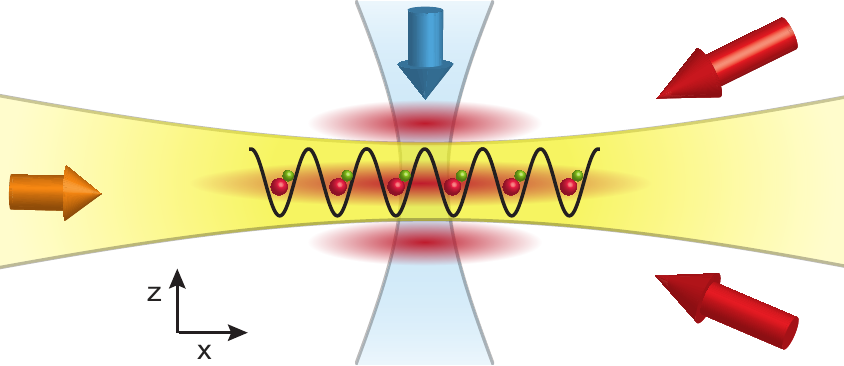}
\caption{\label{fig:apparatus} Schematic of relevant optical potentials.  The yellow beam depicts the 1064\,nm light sheet, and the red beams represent the vertical lattice formed by a 16$^\circ$ intersection of two 1064\,nm beams.  Two overlapping dimple beams with wavelengths 840\,nm and 640\,nm (shown in blue) help load a small, reproducible number of Rb and Na atoms on every experimental cycle.  The 2D lattice at 1064\,nm forms an in-plane sinusoidal potential.}
\end{figure}

At the end of the evaporation, we achieve a dual BEC with $2\times 10^5$ atoms of each species. We perform an RF Landau-Zener transfer of the atoms in both species to the $\left|1,1\right\rangle$ state, the entrance channel for the Feshbach resonance we use. Next, we increase the depth of the light sheet to 2\,W to reduce the size of the clouds and transfer the atoms into a single layer of the vertical lattice (Fig.~\ref{fig:apparatus}).

\section{Bichromatic dimple trap} 

\begin{figure}[tb]
\includegraphics[width=.5\columnwidth]{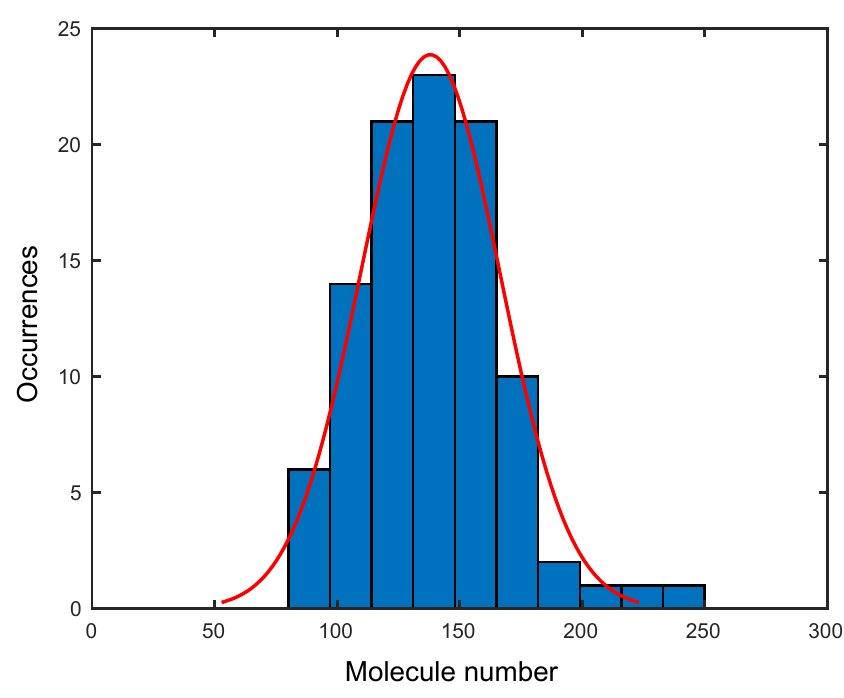}
\caption{\label{fig:numberhist} Example of molecule number distribution achieved through evaporation in the bichromatic dimple trap. Here, the average number of molecules is 138 with a standard deviation of 28.}
\end{figure}

The need to prepare gases with central densities on the order of one particle per site, combined with the spacing of our 2D lattice and the transverse confinement of the vertical lattice, determines the typical atom numbers we need to work with. These turn out to be on the order of a few hundred atoms of each species. Preparing such gases by evaporation in the XODT is unrealistic because of its very weak confinement. Instead, we load a small fraction of the atoms in the dual BEC into a bichromatic dimple trap, with the vertical confinement still provided by the vertical lattice. The dimple trap is formed by two overlapping beams at 840\,nm and 640\,nm, each focused through the microscope objective to a waist of 8\,$\mu$m. We use a 14\,G/cm magnetic field gradient to spill atoms from the reservoir. At the end of the evaporation we lower the dimple depths to zero to load the atoms back into the vertical lattice. By tuning the relative laser powers in the 840\,nm versus 640\,nm dimples (both on the order of 100\,$\mu$W), we can independently control the number of Na and Rb atoms that remain after the evaporation. An example histogram of the molecule number achieved in this way is shown in Fig.~\ref{fig:numberhist}. 

\section{Molecule creation efficiency}

The main limitation on our molecule creation efficiency is the probability to have exactly one atom of each species on a lattice site. In addition, both atoms have to be in the relative motional ground state of the site. Near zero magnetic field, we can create Rb Mott insulators with $\sim85\%$ filling (in the absence of Na). For the mixture, the typical filling we achieve is lower as reported in the main text. It is difficult to characterize the Na filling in the lattice due to low signal-to-noise in absorption imaging. Paths for potential improvement include: (1) increasing the confinement of the vertical lattice to increase the intraspecies Hubbard interaction strengths, (2) optimizing the magnetic field and 2D lattice ramps to reduce the temperature of the mixture, and (3) increasing the overlap between the two species.

For sites that do end up with one atom of each species, the probability of forming a molecule after the ramp through the Feshbach resonance is given by the Landau-Zener formula $p=1-e^{-2\pi\delta_\text{LZ}}$, where $\delta_\text{LZ}=\frac{h n_2}{\mu} \left|\frac{a_\text{bg}\Delta}{\dot{B}}\right|$~\cite{chin2010feshbach,zhang2020forming}. Here, $\mu$ is the reduced mass of the two atoms, $a_\text{bg}=76.33\,a_0$ is the interspecies background scattering length~\cite{guo2021tunable}, $\Delta = 4.255$\,G is the width of the resonance, $\dot{B}=2.5$\,G/ms is the field ramp rate, and $n_2=\int\int n_\text{Na}(\textbf{r})n_\text{Rb}(\textbf{r})d\textbf{r}$ is the pair density on a site. The wavefunctions of the non-interacting atoms in the ground state of a site of a deep lattice can be approximated by the ground state of an anisotropic harmonic oscillator with trap frequencies $\omega_{r,\text{Na}} = 2\pi\times 70.5$\,kHz and $\omega_{z,\text{Na}}=2\pi\times 2.4$\,kHz ($\omega_\text{Rb}=0.86\,\omega_\text{Na}$), which gives $n_2=4.5\times10^{19}$\,m$^{-3}$. We therefore expect $p \approx 1$ for our experimental parameters.

Once we convert the atoms to molecules on these sites, we remove free atoms on the remaining sites using the procedure described in the main text. The four resonant optical removal pulses for each species are 50\,$\mu$s in duration at 1\,mW/cm$^2$. The molecule lifetime in the presence of the Na optical removal light is 1710(120)\,$\mu$s, and the lifetime is 5720(360)\,$\mu$s in the presence of the Rb light. From these lifetimes, we estimate that the free atom removal pulses lead to a loss of $14$\% of the molecules.

\section{Quantum Gas Microscopy}

Here we describe in more detail the process for detecting the Rb atoms that tag the positions of the molecules in the lattice. After breaking apart the molecules and removing the residual Na atoms, the magnetic field is ramped to zero in preparation for laser cooling. The 2D lattice is ramped to a depth of 6000\,$E_{\text{r,Rb}}$ to suppress hopping of the Rb atoms during fluorescence imaging. The light sheet potential is then pinned to a depth of 140\,$\mu$K after which the vertical lattice depth is ramped down to zero.

\begin{figure}[tb]
\includegraphics[width=.5\columnwidth]{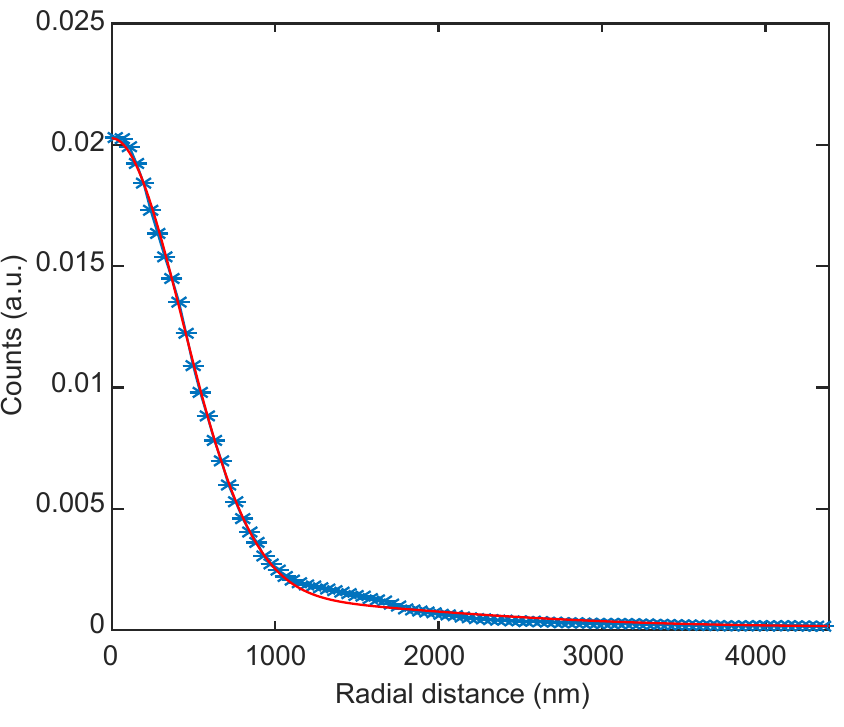}
\caption{\label{fig:psf} Point spread function with radial averaging (blue stars) and a double Gaussian fit (red line).}
\end{figure}

We image the Rb atoms in the lattice by applying optical molasses cooling beams and collecting a portion of the scattered photons. We send two beams each with a 500\,$\mu$m waist containing 30\,$\mu$W of cooling and 6\,$\mu$W of repump light through the side of the vacuum chamber, intersecting at a $90^\circ$ angle in the horizontal plane to produce the requisite polarization gradients. We find that by tilting one of the molasses beams at $8^\circ$ with respect to the horizontal we are able to provide sufficient cooling along the vertical direction. The cooling light is 30\,MHz red-detuned from the $F=2 \rightarrow F'=3$ free-space transition, and the repump light is 42\,MHz blue-detuned from the $F=1 \rightarrow F'=2$ transition. The molasses beams are retroreflected with mirrors mounted on piezo chips oscillating at 300-400\,Hz to smooth out interference patterns in the cooling beams during the exposure time. We use a custom microscope objective with NA$\,=0.5$ (Special Optics 54-25-25) mounted directly above the vacuum chamber to collect $\sim10^4$~photons/atom/second on a sCMOS camera (Andor Zyla 4.2) with 30x magnification. The full-width at half-maximum of the point spread function is 1010(15)\,nm (Fig. \ref{fig:psf}), which allows us to extract the positions of the Rb atoms with single-site resolution using a reconstruction algorithm~\cite{sherson2010single}. 

We characterize the fidelity of our microscopy by measuring the fraction of Rb atoms that hop between lattice sites or are lost during the 0.5\,s exposure time. We measure a hopping rate per exposure of $0.1\,(1)$\% and a loss rate of $1.7\,(3)$\%. The latter is consistent with the background gas-limited lifetime of Rb atoms in the optical trap.

\section{Molecule Detection Fidelity}
An important condition for measuring correlations between molecules is ensuring that we are detecting all of the molecules (low false negative rate) and that the detected particles are in fact molecules and not spurious Rb atoms (low false positive rate). The procedure for measuring the false negative rate is described in the main text. The false positive rate is measured by forming a molecular gas, removing stray atoms, and then reducing the magnetic field to zero without dissociating the molecules. We measure the number of Rb atoms detected and compare it to an equivalent sequence in which we dissociate the molecules before imaging. After 100~repetitions of each sequence, we find that $0.34\,(5)$\% of the Rb tag atoms in the molecule detection sequence are also present in the sequence in which we do not deliberately detect the molecules. The false positives are most likely due to imperfect removal of residual Rb atoms following the molecule formation.

\section{Light-assisted collisions}

When multiple atoms are confined to the same site of a lattice, light-assisted collisions can cause pairs of atoms to be ejected from the trap when illuminated with resonant light~\cite{fuhrmanek2012light,kaufman2021quantum}. Since our molecule detection procedure requires removing Na atoms with resonant $F=2 \rightarrow F^\prime = 3$ light, it is important to determine if light-assisted collisions during the optical pulse can also remove the Rb atoms which tag the molecule positions. We estimate that the energy gained per photon from a light-assisted collision is about fourteen orders of magnitude smaller for Na-Rb pairs than it is for Rb-Rb pairs and is therefore negligible in our system. 

The large difference in energy is due to the fact that for homonuclear light-assisted collisions the atoms experience an attractive potential that scales with the internuclear distance $R$ as $1/{R^3}$, while for heteronuclear collisions the interaction potential scales as $1/{R^6}$. This means that a heteronuclear atom pair gains significantly less energy during each scattering event than for a homonuclear pair. Using known values for the $C_3$ and $C_6$ coefficients \cite{marinescu1995dispersion,marinescu1999long}, we estimate the energy gained by the atoms using a classical toy model in the center-of-mass frame with reduced mass $\mu$. We assume that the atoms are initially at rest with a separation of approximately 100\,nm on a lattice site and that the time spent in the excited state is given by $\tau = 1/\Gamma$, where $\Gamma = 2\pi \times 10$\,MHz ($2\pi \times 6$\,MHz) is the decay rate for Na (Rb). We then numerically integrate Newton's equation $d^2R/dt^2 = -3C_3/\mu R^4$ for the case of two Rb atoms, or $d^2R/dt^2 = -6C_6/\mu R^7$ for the case of one Rb and one Na atom, to determine the change in the internuclear separation during the excited state lifetime and hence the energy gained per scattering event.

\section{Molecule lifetimes}

\begin{figure}[tb]
\includegraphics[width=\columnwidth]{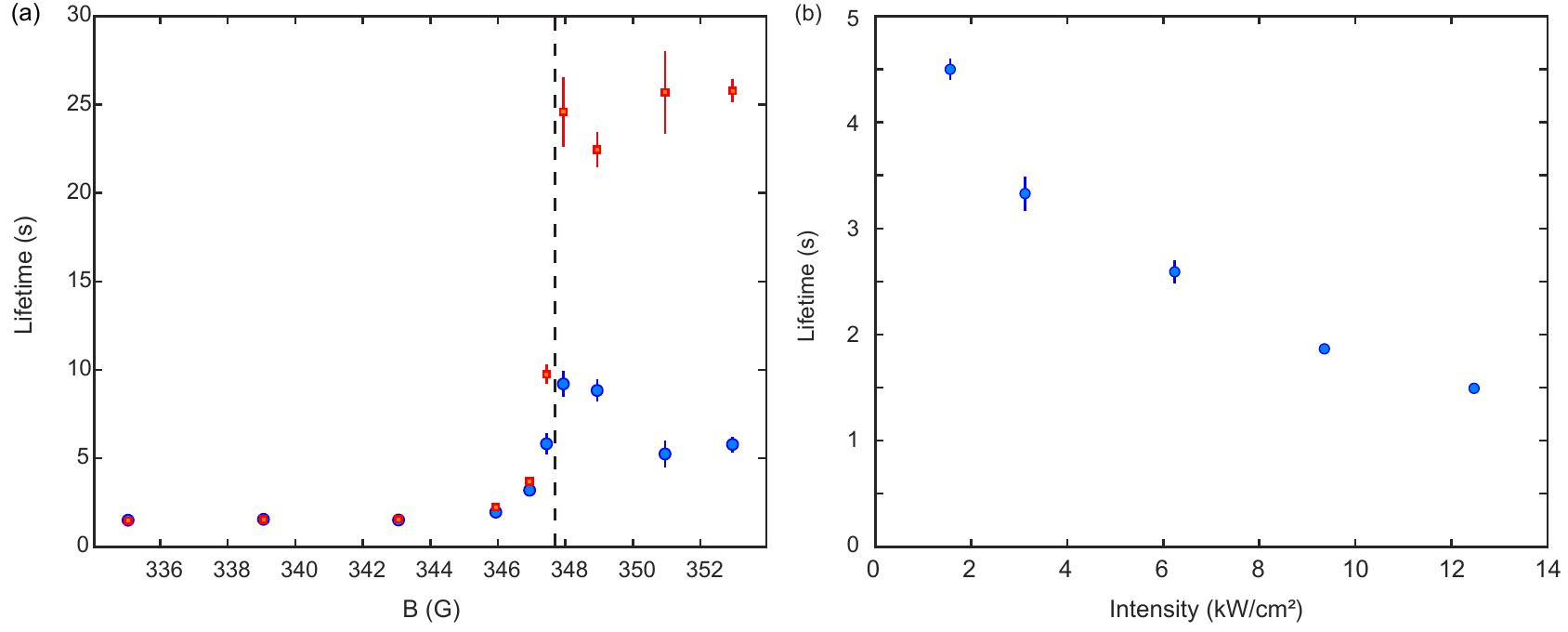}
\caption{\label{fig:lifetimes} Molecule lifetime in an optical lattice. (a) Molecule lifetime versus magnetic field for a lattice intensity of 12.5\,kW/cm$^2$. Blue circles indicate the pure molecular lifetime where free atoms are removed both before and after the hold time, while the red squares do not include the second free atom removal. (b) Lifetime versus 2D lattice depth at 335.1\,G.}
\end{figure}

We measure the lifetime of the NaRb molecules as a function of lattice depth and magnetic field. As shown in Fig.~\hyperref[fig:lifetimes]{S4(a)}, performed at our typical 2D lattice intensity of 12.5\,kW/cm$^2$, the molecule lifetime in the lattice is sufficiently long at all magnetic fields studied that loss can be ignored for the experiments presented in this paper. Following the removal of residual unassociated atoms at 346.6\,G, we ramp the magnetic field to the target value and hold for a variable length of time. We then return the magnetic field to 346.6\,G and repeat the removal of free atoms before dissociating and imaging the surviving molecules (blue circles).

On the repulsive side of the Feshbach resonance, the molecule lifetime initially decreases with decreasing magnetic field before saturating at $\sim1.5$\,s. This follows the magnetic field dependence of the closed channel fraction of the Feshbach molecules. As has been previously observed, larger closed channel fractions increase the wavefunction overlap with electronically excited molecular states, enhancing molecule loss due to off-resonant excitation by the lattice beams~\cite{danzl2009deeply,chotia2012long}. Interestingly, we find that the lifetime of the molecules also decreases above the Feshbach resonance. Since the molecules are only weakly bound due to confinement effects above the resonance~\cite{busch1998two,stoferle2006molecules}, we explore whether at these fields the molecule lifetime is limited by a mechanism that breaks apart the molecules but does not eject the remaining atoms from the trap, such as noise in the magnetic field. We therefore repeat this measurement without the final removal of stray atoms (red squares). The molecule lifetimes show the same behavior as before on the repulsive side of the resonance, but on the attractive side we now see long lifetimes consistent with background gas-limited atomic lifetimes. This indicates that dissociation of the molecules does in fact occur in our system when holding the weakly bound molecules above the Feshbach resonance.

To examine the role of the light scattering on these lifetimes, we also measure the lifetimes at a fixed field of 335.1\,G for five different lattice intensities (Fig.~\hyperref[fig:lifetimes]{S4(b)}). We see that the lifetime increases for shallower lattice depths, consistent with reduced off-resonant excitation of the molecules by the lattice. For much shallower lattice depths (not shown), molecule tunneling occurs and a single-body loss rate is no longer appropriate due to inelastic collisions between the molecules.

\section{Binding energy estimation}
To obtain each point in the dissociation spectrum, we fix the microwave frequency and scan the magnetic field by $\sim$\,130\,mG, corresponding to a $\sim\,300$\,kHz Landau-Zener sweep. The molecular binding energies reported in Fig.~3 of the main text are extracted by fitting the dissociation curves to a phenomenological function $N(\omega)$, where $N$ is the number of Na atoms transferred from the molecular state to the free $|2,2\rangle$ state:
\begin{align}
    N(\omega) &= \frac{p_1}{f(\omega)} e^{-8\log(2) ((x-p_2)/f(\omega))^2}  \\
    f(\omega) &= \frac{2p_3}{1+e^{p_4(\omega-p_2)}}\nonumber
\end{align}
Here, $p_i$ are free fitting parameters. The binding energy is estimated as the frequency at which the transfer $N(\omega)$ is half the peak of the fitted lineshape.

\end{document}